\documentclass{PoS}

\usepackage{graphicx}
\usepackage{bm}
\usepackage{epsfig}
\usepackage{graphics}
\usepackage{amsmath,amsfonts,amssymb}

\def\Dsl{\hbox{/\kern-.6000em D}} 

\def\dsl{\,\raise.15ex\hbox{/}\mkern-13.5mu D}
\def\bsigma{\mbox{\boldmath $\sigma$}}

\def\bsigma{\mbox{\boldmath $\sigma$}}

\def\ltap{\ \raise.3ex\hbox{$<$\kern-.75em\lower1ex\hbox{$\sim$}}\ }
\def\gtap{\ \raise.3ex\hbox{$>$\kern-.75em\lower1ex\hbox{$\sim$}}\ }
\def\OMIT#1{}

\def\lsim{\mathrel{\raise.3ex\hbox{$<$\kern-.75em\lower1ex\hbox{$\sim$}}}}
\def\gsim{\mathrel{\raise.3ex\hbox{$>$\kern-.75em\lower1ex\hbox{$\sim$}}}}

\def\msb{{\overline{\rm MS}}}

\newcommand{\nn}{\nonumber}

\newcommand{\bmp}{\mathbf p}

\newcommand{\bmJ}{\mathbf J}

\newcommand{\bmpp}{{\bmp^\prime}}

\newcommand{\ord}{{\cal O}}

\def\msb{{\overline{\rm MS}}}

\catcode`\@=11
\def\slash{\mathpalette\make@slash}
\def\make@slash#1#2{\setbox\z@\hbox{$#1#2$}%
  \hbox to 0pt{\hss$#1/$\hss\kern-\wd0}\box0}
\catcode`\@=12 

\newcommand{\be}{\begin{equation}}
\newcommand{\ee}{\end{equation}}
\newcommand{\bea}{\begin{eqnarray}}
\newcommand{\eea}{\end{eqnarray}}

\title{NNLL top-antitop production at threshold\thanks{preprint UWThPh-2011-40}}

\ShortTitle{NNLL top-antitop production at threshold}

\author{\speaker{Maximilian Stahlhofen}\\
        University of Vienna, Faculty of Physics, Boltzmanngasse 5, A-1090 Wien, Austria\\
        E-mail: \email{Maximilian.Stahlhofen@univie.ac.at}}

\author{Andr\'e Hoang\\
        University of Vienna, Faculty of Physics, Boltzmanngasse 5, A-1090 Wien, Austria\\
        E-mail: \email{Andre.Hoang@univie.ac.at}}

\abstract{We present an update to the NNLL RG-improved QCD prediction of top-antitop production in $e^+ e^-$ annihilation at threshold. It includes for the first time a complete NNLL resummation of ultrasoft logarithms, which are dominant at this order and give a sizable correction. The renormalization scale dependence of the total resonance cross section decreases substantially compared to earlier predictions, where the ultrasoft logarithms were included only partially.}

\FullConference{10th International Symposium on Radiative Corrections (Applications of Quantum Field Theory to Phenomenology)\\
		 September 26-30, 2011\\
		 Mamallapuram, India}

\begin{document}

\section{Introduction}

A scan of the resonance in top-antitop production at a future linear collider will allow for a precise determination of the mass, the width and the couplings of the top quark~\cite{thresholdscan}. This imposes high requirements on the precision of the theoretical prediction for the total cross section near threshold, which can in principle be met by high order perturbative calculations using nonrelativistic effective field theories based on NRQCD~\cite{BBL}. So far, however the accuracy in the normalization of the total $t\,\bar t$ resonance cross section was limited by a relatively large renormalization scale dependence of about 6\,\% in the renormalization group improved (RGI) NNLL prediction~\cite{HoangEpiphany} (see also Ref.~\cite{PinedaSigner}) and about 10\,\% in a fixed order N$^3$LO computation~\cite{Beneke:2008ec}. Although the determination of the top mass, which primarily depends on the c.m. energy where the cross section rises, might not be affected by this uncertainty, it still renders precision measurements of the top total width or the top Yukawa coupling impossible. To match the statistical uncertainties, that are expected for these quantities at the International Linear Collider (ILC), a theoretical precision of the cross section normalization of at least 3\% is needed. 

In this talk we shortly review recent progress in the RGI calculations concerned with heavy quark-antiquark production at threshold within the vNRQCD effective theory framework~\cite{LMR} and discuss its implications for the resonance line-shape of $\sigma_{\rm tot}(e^+ e^- \to t\,{\bar t})$.

\section{The cross section}

To achieve reliable results for a heavy quark pair production process close to threshold it is mandatory to resum so-called ``Coulomb singularities'', i.e. terms that scale like powers of $\alpha_s/v$, where $\alpha_s$ is the strong coupling and $v$ is the (nonrelativistic) relative velocity of the heavy quarks, to all orders in perturbation theory. This task can be performed employing a Schr\"odinger equation within the nonrelativistic effective theory approach. In addition potentially large logarithms of the relative velocity $v$ as they typically appear in quantum corrections to the leading order cross section can be systematically resummed using a modified renormalization group with the ``subtraction velocity'' $\nu$ being the renormalization scale parameter~\cite{LMR}. Thus the R-ratio for top-antitop threshold production schematically takes the form~\cite{HMST}
\begin{align}
R = \frac{\sigma_{t \bar t}}{\sigma_{\mu^+\mu^-}} = v \sum\limits_k \bigg(\frac{\alpha_s}{v}\bigg)^k \sum\limits_i (\alpha_s\,\ln\, v )^i \times \bigg\{1 \, ({\rm LL});\;\alpha_s, v \, ({\rm NLL});\;\alpha_s^2,\, \alpha_s v,\, v^2\,({\rm NNLL});\;\ldots \bigg\}
\label{Rstruc}
\end{align}
and we assume $v\sim\alpha_s$ due to the Coulombic bound-state-like dynamics of the system in the resonance region.

The R-Ratio for the production via a virtual photon or Z boson with the c.m. energy $\sqrt{s}$ has contributions from vector and axial-vector currents and can be written as
\begin{align}
R^{\gamma,Z}(s) = F^v(s)\,R^v(s) +  F^a(s) R^a(s) \,, \label{Rtot}
\end{align}
where due to the optical theorem
\begin{align}
 R^v(s) &= \frac{4 \pi }{s}\,\mbox{Im}\,\left[-i\int d^4x\: e^{i\sqrt{s}\,t }
  \left\langle\,0\,\left|\, T\, j^v_{\mu}(x) \,
  {j^v}^{\mu} (0)\, \right|\,0\,\right\rangle\,\right] \,, \nn\\
 R^a(s) &= \frac{4 \pi }{s}\,\mbox{Im}\,\left[-i\int d^4x\: e^{i\sqrt{s}\,t }
  \left\langle\,0\,\left|\, T\, j^a_{\mu}(x) \,
  {j^a}^{\mu} (0)\, \right|\,0\,\right\rangle\,\right] \,.
\end{align}
The prefactors $F^v(s)$, $F^a(s)$ contain the effects from the $\gamma$ and Z exchange and are given in Ref.~\cite{HMST}. The standard model currents $j^v_{\mu}$ and $j^a_{\mu}$ produce the heavy quark pair in a vector and an axial-vector state, respectively. Within the effective theory description these currents are replaced by their nonrelativistic counterparts and we find to NNLO in the $v$ counting
\begin{align}
R^v(s) &= \frac{4\pi}{s}\,
 \mbox{Im}\Big[\,
 c_1^2(\nu)\,{\cal A}_1(v,m,\nu) + 
 2\,c_1(\nu)\,c_2(\nu)\,{\cal A}_2(v,m,\nu) \,\Big] \,, 
\nn\\
 R^a(s) &=  \frac{4\pi}{s}\,
 \mbox{Im}\Big[\,c_3^2(\nu)\,{\cal A}_3(v,m,\nu)\,\Big]\,,
\label{effRratios}
\end{align}
where the ${\cal A}_i$ denote effective current correlators and ${\cal A}_2$ and ${\cal A}_3$ are suppressed compared to the LO correlator
\begin{align}
{\cal A}_1(v, m, \nu) &= i\,
 \sum\limits_{\bmp,\bmpp}
 \int\! d^4x\: e^{i (\sqrt{s} - 2 m) t}\:
 \Big\langle\,0\,\Big|\, T\, \bmJ_{1,\bmp}(x) \, \bmJ^\dagger_{1,\bmpp}(0)
 \Big|\,0\,\Big\rangle 
\label{A1}
\end{align}
by a factor of $v^2$. The $c_i \sim 1$ are Wilson coefficients of the effective currents, i.e. functions of the renormalization scale $\nu$ and the heavy quark mass $m$.
In Eq.~\eqref{A1} we have adopted the vNRQCD label notation for the leading ${}^3S_1$ current
\begin{align} 
  {\bmJ}_{1,\bmp} = 
    \psi_{\bmp}^\dagger\, \bsigma (i\sigma_2) \chi_{-\bmp}^*
\,,
\label{J1J0}
\end{align}
where  $\psi_{\bmp}$ and $\chi_{\bmp}$ annihilate top and antitop quarks with
three-momentum $\bmp$, respectively, and where color indices have been suppressed.

The missing piece to reach NNLL precision in Eq.~\eqref{effRratios} is the NNLL running of the S-Wave current coefficient $c_1(\nu)$. All other pieces including all relevant electroweak effects\footnote{Concerning electroweak effects we refer to cross section predictions with moderate cuts on the reconstructed top invariant mass~\cite{Hoang:2010gu}. See Ref.~\cite{Beneke:2010mp} for results with more general invariant mass cuts.} ($\alpha_{ew} \sim \alpha_s^2$) are known with sufficient accuracy~\cite{HMST,PinedaSigner,Hoang:2010gu,Beneke:2010mp,Penin:2011gg}. In particular, the ${\cal A}_i$ correlators can be related to Green functions of a Schr\"odinger operator and incorporate the resummation of Coulomb singular terms.

\section{Current renormalization}

Without loss of generality we parameterize the RG evolution of the current coefficient $c_1(\nu)$ as
\begin{align}
 \ln\Big[ \frac{c_1(\nu)}{c_1(1)} \Big] & =
\xi^{\rm NLL}(\nu) + 
\Big(\,
\xi^{\rm NNLL}_{\rm m}(\nu) + \xi^{\rm NNLL}_{\rm nm}(\nu)
\,\Big) + \ldots
\,,
\label{c1solution}
\end{align}
where the LL contribution vanishes~\cite{LMR} and the NNLL contribution consists of a ``mixing'' term $\xi^{\rm NNLL}_{\rm m}$ and a ``non-mixing'' term $\xi^{\rm NNLL}_{\rm nm}$. The latter is generated directly by UV divergent three-loop vertex diagrams for the production of a heavy quark-antiquark pair and has been computed in Ref.~\cite{3loop} using vNRQCD.
It involves a non-trivial entanglement of soft and ultrasoft scales at subleading order that has at this time only been systematically treated in vNRQCD.
 The ``mixing'' part, in contrast, is generated by the same two-loop diagrams that are responsibe for the NLL running of $c_1$ through subleading NLL corrections to the (four-quark) vertices, namely the potentials, in the loops.

In other words, we can derive $\xi^{\rm NNLL}_{\rm m}$ from the known NLL anomalous dimension~\cite{ManoharVk,Pineda:2001et,HoangStewartultra} (${\bf S}^2=2$)
\begin{align}
 \left(\nu \frac{\partial}{\partial\nu} \ln[c_1(\nu)] \right)^{\rm NLL} 
\!\!\!\! =  
 -\:\frac{{\cal V}_c^{(s)}(\nu)
  }{ 16\pi^2} \bigg[ \frac{ {\cal V}_c^{(s)}(\nu) }{4 }
  +{\cal V}_2^{(s)}(\nu)+{\cal V}_r^{(s)}(\nu)
   + {\bf S}^2\: {\cal V}_s^{(s)}(\nu)  \bigg] 
 +\frac12 {\cal V}_{k,\rm eff}^{(s)}(\nu)
\label{c1anomdim}
\end{align}
through the NLL matching and RG evolution corrections to the potential coefficients ${\cal V}_i^{(s)}$~\cite{ManoharVk,amis}.
Unfortunately the determination of the subleading NLL running of the vNRQCD potentials is not yet complete except for the Coulomb potential ${\cal V}_c^{(s)}$~\cite{Pineda:2001ra,HoangStewartultra} and the spin-dependent potential ${\cal V}_s^{(s)}$~\cite{Penin:2004xi}.
Recently however we have finished the calculation of the dominant, ultrasoft contributions to the NLL running of the remaining potential coefficients ${\cal V}_2^{(s)}$, ${\cal V}_r^{(s)}$ and ${\cal V}_{k,\rm eff}^{(s)}$ relevant in Eq.~\eqref{c1anomdim}. In that work~\cite{V2Vr,Vk} we have determined these contributions from vNRQCD diagrams with four external heavy quark legs and two ultrasoft gluon loops. Ultrasoft gluons carry energy and momentum of $\ord(mv^2)$ and are to be distinguished from soft degrees of freedom with energy and momentum of $\ord(mv)$. The same results were independently obtained also in a different effective theory framework called pNRQCD~\cite{Pineda:1997bj} in Ref.~\cite{Pineda:2011aw} and allowed us to determine the full ultrasoft contribution to the NNLL mixing term $\xi^{\rm NNLL}_{\rm m}$ in Ref.~\cite{Vk}.

The analysis of the three-loop (non-mixing) terms in Ref.~\cite{3loop} showed
that the contributions involving the exchange of ultrasoft gluons are more than an order of magnitude
larger than those arising from soft matrix element insertions and in fact
similar in size to the NLL contributions. The reason is related to the larger size of the ultrasoft coupling $\alpha_s(m \nu^2)$ compared to the soft coupling $\alpha_s(m \nu)$ and to a rather large coefficient multiplying the ultrasoft contributions. 
From this analysis it is reasonable to assume that the ultrasoft effects, which form a gauge-invariant subset, also
dominate the mixing contributions. This is also consistent with the small
numerical effects~\cite{Penin:2004ay} of the NLL evolution of the spin-dependent
coefficient ${\cal V}_s^{(s)}$~\cite{Penin:2004xi}, which is dominated by soft
effects and receives ultrasoft contributions only indirectly through mixing.

The ultrasoft contribution to $\xi^{\rm NNLL}_{\rm m}$ determined in Ref.~\cite{Vk} reads\footnote{The form of Eq.~\eqref{XiNNLLmix} implies the convention that the $z$ and $\omega=1/(2-z)$ parameters in $\xi^{\rm NLL}$, as given in Ref.~\cite{HoangStewartultra}, are according to Eq.~\eqref{zLL}.}
\begin{align}
\xi^{\rm NNLL}_{\rm m, usoft} & = 
\frac{2\pi \beta_1}{\beta_0^3}\,\tilde A\,\alpha_s^2(m)\,
 \bigg[ -\frac{7}{4}+\frac{\pi^2}{6}+z\left(1-\ln\frac{z}{2-z}\right)
      +z^2\left(\frac{3}{4}-\frac{1}{2}\ln z\right)
\nn\\[1.5 ex] & \hspace{16 ex} 
      -\ln^2\left(\frac{z}{2}\right)+\ln^2\left(\frac{z}{2-z}\right)
      -2\mbox{Li}_2\left(\frac{z}{2}\right)\bigg]
\nn\\[1.5 ex] & 
+\,\frac{8\pi^2}{\beta_0^2}\,\tilde B\,\alpha_s^2(m)\,
  \Big[ 3-2z-z^2-4\ln(2-z) \Big]
\,,
\label{XiNNLLmix}
\end{align}
with
\begin{align}
\beta_0 &= \frac{11}{3}C_A - \frac43 T n_f \,,\quad \beta_1 = \frac{34}{3}C_A^2 - 4 C_F T n_f - \frac{20}{3}C_A T n_f\,,\\
\tilde A &= -C_F(C_A+C_F)(C_A+2C_F) \frac1{3\pi}
 \,, \label{ASchlange} \\
\tilde B &= -C_F(C_A+C_F)(C_A+2C_F) \frac{C_A(47 + 6\pi^2) -10 n_f T }{108 \pi ^2} \,, \label{BSchlange} 
\\
z & \equiv  \left(\frac{\alpha_s(m\nu)}{\alpha_s(m)}\right)^{\rm LL}
  \, = \,
  \bigg(1+\frac{\alpha_s(m)\beta_0}{2\pi}\ln\nu\bigg)^{-1}\,. 
\label{zLL}
\end{align}
Together with $\xi^{\rm NNLL}_{\rm nm}$ from Ref.~\cite{3loop} it gives the dominant contribution to the NNLL running of $c_1(\nu)$ (in the $\msb$ scheme). The corresponding matching condition $c_1(1)$ in Eq.~\eqref{c1solution} is obtained from a full theory two-loop calculation~\cite{Czarnecki:1997vz,Beneke:1997jm} and is given in Ref.~\cite{3loop}.

\section{Results}

\begin{figure}[ht]
\begin{center}
\includegraphics[width=0.95\textwidth]{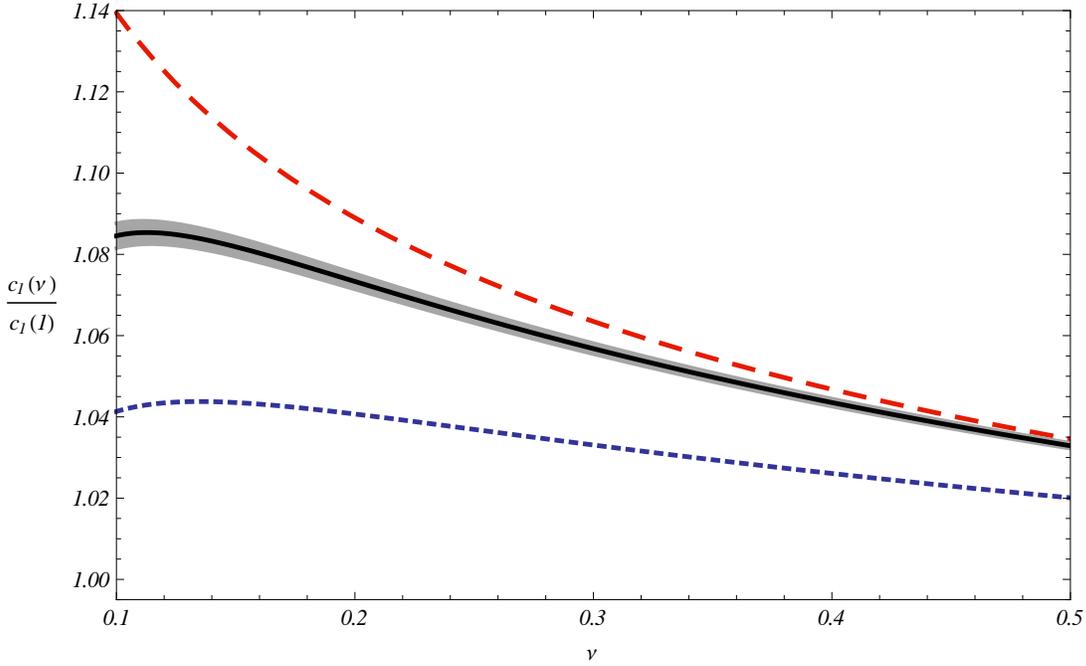}
\end{center}
\caption{RG evolution of the  ${}^3S_1$ current coefficient
$c_1(\nu)$ normalized to $c_1(1)$ for top-antitop production ($m=175$~GeV). The
dotted (blue) line represents the full NLL result 
$\exp[\xi^{\rm NLL}]$, the dashed (red) line includes in addition the NNLL
non-mixing contribution, $\exp[\xi^{\rm NLL} + \xi_{\rm nm}^{\rm NNLL}]$. The
solid (black) line accounts for the full NLL and NNLL non-mixing contributions as
well as for the new ultrasoft NNLL mixing corrections, $\exp[\xi^{\rm NLL} +
\xi_{\rm nm}^{\rm NNLL} +\xi^{\rm NNLL}_{\rm m, usoft}]$. The gray area around the black line is generated by varying the soft NNLL non-mixing contributions to that curve by a factor between 0 and 2.
For the plot we have used $\alpha_s^{(n_f=5)}(175\;\mbox{GeV})=0.107$. \label{c1Plot}} 
\end{figure}

The numerical analysis of the new result for the NNLL evolution of the current coefficient $c_1(\nu)$ carried out in Ref.~\cite{Vk} shows that the ultrasoft mixing contribution in Eq.~\eqref{XiNNLLmix} compensates the anomalously large ultrasoft nonmixing contribution to a large extent. That can be observed best in the plot of Fig.~\ref{c1Plot}. The renormalization scale dependence of $c_1(\nu)$ reduces substantially, when the new result in Eq.~\eqref{XiNNLLmix} is included in the NNLL prediction. This is especially so close to the resonance, where typically $\nu \sim v \sim 0.15$.

To illustrate the (small) effect of the known soft contributions at NNLL we have drawn the gray area around our new NNLL result (black solid line) in Fig.~\ref{c1Plot}. It shows the variation of the result due to multiplying the soft non-mixing terms at NNLL order by a factor between zero and two. 
We believe that this variation represents a good estimate of the uncertainty associated to the currently unknown NNLL soft mixing contributions. Compared to the remaining scale dependence of the NNLL ultrasoft contributions this uncertainty is negligible.

\begin{figure}[t]
\includegraphics[width=\textwidth]{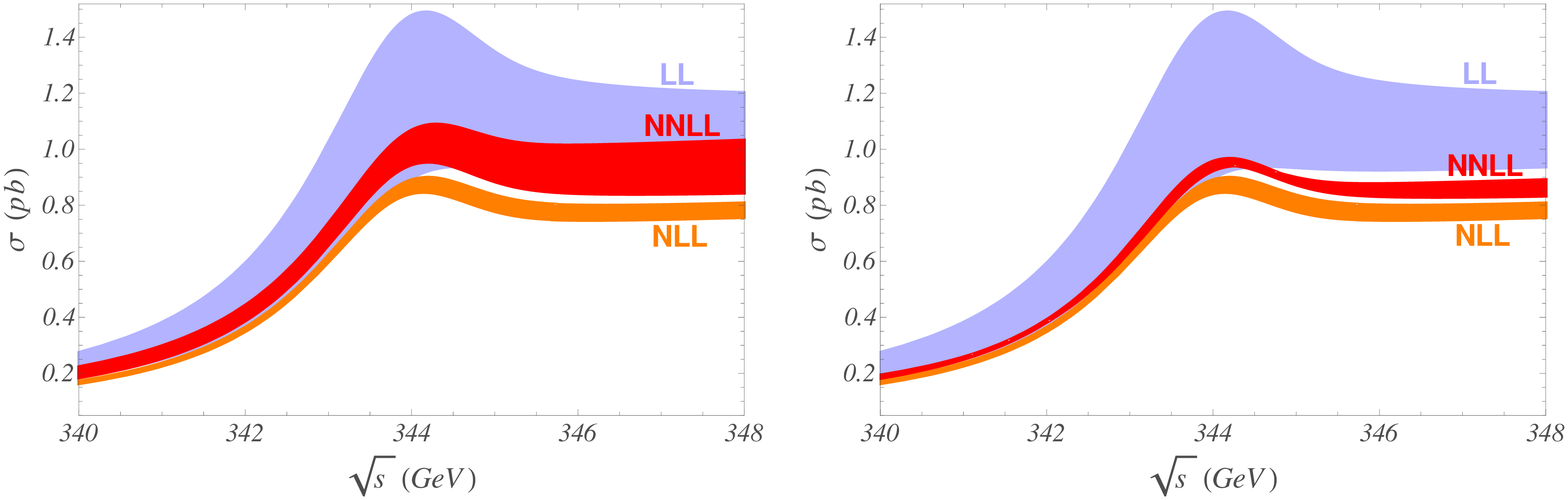}
\put(-395,120){a)}
\put(-180,120){b)}
\caption{
The band labeled ``NNLL'' represents the prediction of $\sigma_{\rm tot}(e^+ e^- \to t\,{\bar t})$ in the threshold region with (a) and without (b) the NNLL ultrasoft mixing contribution to $c_1(\nu)$ in addition to all other known QCD corrections up to NNLL order. The other bands refer to the complete NLL and LL results, respectively and are identical in both panels. All bands were generated by varying the renormalization parameter $\nu$ between $0.1$ and $0.4$ and using the 1S-mass scheme~\cite{Hoang:1999zc} with $m^{1S}=172\;\mbox{GeV}$. Further input parameters were $\Gamma_t=1.46\;\mbox{GeV}$ for the top width and $\alpha_s(M_Z)=0.118$ for the strong coupling in the $\msb$ scheme.
In the peak region of the NNLL result in panel b the scale variation is around $\pm 1.7 \%$.
\label{CrossSections}} 
\end{figure}

Finally we would like to show the effect of the new NNLL result in Eq.~\eqref{XiNNLLmix} on the prediction of $\sigma_{\rm tot}(e^+ e^- \to t\,{\bar t})$ in the resonance region. In Fig.~\ref{CrossSections} a we have plotted the threshold cross section at LL, NLL and NNLL order without the new NNLL ultrasoft mixing contribution to the current as it was presented in Ref.~\cite{HoangEpiphany}. In Fig.~\ref{CrossSections} b the cross section is shown including the new NNLL ultrasoft mixing contribution.\footnote{For both predictions the corresponding Schr\"odinger equation with the NNLL Coulomb potential was solved ``exactly'' using numerical methods. Relativistic corrections were consistently treated as perturbations.} The decay of the top, actually an electroweak effect, we implemented only at LO by shifting the c.m. energy $\sqrt{s} \to \sqrt{s} - i\Gamma_t$. A systematic treatment of electroweak effects up to NNLL order can be found in Refs.~\cite{Hoang:2010gu,Hoang:2006pd,Hoang:2004tg,Beneke:2010mp,Penin:2011gg}.

At any order the velocity renormalization parameter $\nu$ has been varied between 0.1 and 0.4. The reduction of the renormalization scale dependence due to the new NNLL contribution is substantial and clearly visible in Fig.~\ref{CrossSections} b. However it is conspicuous that the gap between NLL and NNLL bands is essentially the same in Fig.~\ref{CrossSections} a and b and that the remaining NNLL scale variation in Fig.~\ref{CrossSections} b is  comparable to the one of the NLL result.
We believe that this is related to an anomalously small scale dependence of the NLL prediction and might not be interpreted as an instability. We postpone the determination of a final number for the remaining theoretical uncertainty of the NNLL cross section to an upcoming publication.

\section{Conclusions}

Including the recently obtained ultrasoft NNLL mixing contribution to the running of the leading vector current in the prediction for the total production cross section of a top-antitop pair near threshold leads to a substantial reduction of the renormalization scale dependence. We argue that the uncertainty due to the still unknown NNLL soft mixing contribution to the running of the effective current, which is now the only missing piece to a complete NNLL prediction of $\sigma_{\rm tot}(e^+ e^- \to t\,{\bar t})$, can be neglected compared to the size of the now fully known ultrasoft contributions. These findings will result in a substantial reduction of the perturbative uncertainty of the NNLL prediction.
A thorough analysis of the respective theoretical errors from all possible sources including electroweak corrections and a variation of the vNRQCD matching scale in addition to the usual variation of the renormalization scale will be the subject of a future publication.


\begin{thebibliography}{99}


\bibitem{thresholdscan}
  J.~A.~Aguilar-Saavedra {\it et al.}  [ECFA/DESY LC Physics Working Group],
  arXiv:hep-ph/0106315;
  T.~Abe {\it et al.}  [American Linear Collider Working Group],
  in {\it Proc. of the APS/DPF/DPB Summer Study on the Future of
  Particle Physics (Snowmass 2001) } ed. N.~Graf, 
  arXiv:hep-ex/0106057;
  A.~Juste {\it et al.},
  arXiv:hep-ph/0601112.


\bibitem{BBL}
  G.~T.~Bodwin, E.~Braaten and G.~P.~Lepage,
  Phys.\ Rev.\ D {\bf 51}, 1125 (1995)
  [Erratum-ibid.\ D {\bf 55}, 5853 (1997)]
  [arXiv:hep-ph/9407339].


\bibitem{HoangEpiphany}
  A.~H.~Hoang,
  Acta Phys.\ Polon.\ B {\bf 34}, 4491 (2003)
  [arXiv:hep-ph/0310301].


\bibitem{PinedaSigner}
  A.~Pineda and A.~Signer,
  arXiv:hep-ph/0607239.


\bibitem{Beneke:2008ec}
  M.~Beneke, Y.~Kiyo, K.~Schuller,
  PoS {\bf RADCOR2007}, 051 (2007).
  [arXiv:0801.3464 [hep-ph]].


\bibitem{LMR} M.~Luke, A.~Manohar and I.~Rothstein,
Phys.\ Rev.\  {\bf D61}, 074025 (2000)
[arXiv:hep-ph/9910209].



\bibitem{HMST}
  A.~H.~Hoang, A.~V.~Manohar, I.~W.~Stewart and T.~Teubner,
  Phys.\ Rev.\ Lett.\  {\bf 86}, 1951 (2001)
  [arXiv:hep-ph/0011254];
    A.~H.~Hoang, A.~V.~Manohar, I.~W.~Stewart and T.~Teubner,
  Phys.\ Rev.\ D {\bf 65}, 014014 (2002)
  [arXiv:hep-ph/0107144].


\bibitem{Hoang:2010gu}
  A.~H.~Hoang, C.~J.~Reisser, P.~Ruiz-Femenia,
  Phys.\ Rev.\  {\bf D82}, 014005 (2010).
  [arXiv:1002.3223 [hep-ph]].


\bibitem{Beneke:2010mp}
  M.~Beneke, B.~Jantzen, P.~Ruiz-Femenia,
  Nucl.\ Phys.\  {\bf B840}, 186-213 (2010).
  [arXiv:1004.2188 [hep-ph]].

\bibitem{Penin:2011gg}
  A.~A.~Penin, J.~H.~Piclum,
  [arXiv:1110.1970 [hep-ph]].


\bibitem{3loop}
 A.~H.~Hoang,
  Phys.\ Rev.\ D {\bf 69}, 034009 (2004)
  [arXiv:hep-ph/0307376].


\bibitem{ManoharVk}
  A.~V.~Manohar and I.~W.~Stewart,
  Phys.\ Rev.\ D {\bf 63}, 054004 (2001)
  [arXiv:hep-ph/0003107].


\bibitem{Pineda:2001et}
  A.~Pineda,
  Phys.\ Rev.\ D {\bf 66}, 054022 (2002)
  [arXiv:hep-ph/0110216].



\bibitem{HoangStewartultra}
A.~H.~Hoang and I.~W.~Stewart,
Phys.\ Rev.\ D {\bf 67}, 114020 (2003)
[arXiv:hep-ph/0209340].

\bibitem{amis} A.V.~Manohar and I.W.~Stewart,
Phys.\ Rev.\ D {\bf 62}, 014033 (2000)
[arXiv:hep-ph/9912226].


\bibitem{Pineda:2001ra}
  A.~Pineda,
  Phys.\ Rev.\ D {\bf 65}, 074007 (2002)
  [arXiv:hep-ph/0109117];


\bibitem{Penin:2004xi}
  A.~A.~Penin, A.~Pineda, V.~A.~Smirnov and M.~Steinhauser,
  Phys.\ Lett.\ B {\bf 593}, 124 (2004)
  [arXiv:hep-ph/0403080].


\bibitem{V2Vr}
  A.~H.~Hoang and M.~Stahlhofen,
  Phys.\ Rev.\  D {\bf 75}, 054025 (2007)
  [arXiv:hep-ph/0611292].


\bibitem{Vk}
  A.~H.~Hoang, M.~Stahlhofen,
  JHEP {\bf 1106}, 088 (2011).
  [arXiv:1102.0269 [hep-ph]].



\bibitem{Pineda:1997bj}
  A.~Pineda and J.~Soto,
  Nucl.\ Phys.\ Proc.\ Suppl.\  {\bf 64}, 428 (1998).


\bibitem{Pineda:2011aw}
  A.~Pineda,
  Phys.\ Rev.\  {\bf D84}, 014012 (2011).
  [arXiv:1101.3269 [hep-ph]].


\bibitem{Penin:2004ay}
  A.~A.~Penin, A.~Pineda, V.~A.~Smirnov and M.~Steinhauser,
  Nucl.\ Phys.\ B {\bf 699}, 183 (2004)
  [arXiv:hep-ph/0406175].


\bibitem{Czarnecki:1997vz}
  A.~Czarnecki, K.~Melnikov,
  Phys.\ Rev.\ Lett.\  {\bf 80}, 2531-2534 (1998).
  [hep-ph/9712222].

\bibitem{Beneke:1997jm}
  M.~Beneke, A.~Signer, V.~A.~Smirnov,
  Phys.\ Rev.\ Lett.\  {\bf 80}, 2535-2538 (1998).
  [hep-ph/9712302].


\bibitem{Hoang:1999zc}
  A.~H.~Hoang, T.~Teubner,
  Phys.\ Rev.\  {\bf D60}, 114027 (1999).
  [hep-ph/9904468].


\bibitem{Hoang:2006pd}
  A.~H.~Hoang, C.~J.~Reisser,
  Phys.\ Rev.\  {\bf D74}, 034002 (2006).
  [hep-ph/0604104].

\bibitem{Hoang:2004tg}
  A.~H.~Hoang, C.~J.~Reisser,
  Phys.\ Rev.\  {\bf D71}, 074022 (2005).
  [hep-ph/0412258].


\end{thebibliography}
\end{document}